\documentclass[aps,preprint,nofootinbib]{revtex4}
\usepackage{graphicx}
\usepackage{amsmath,amssymb}
\usepackage{url}
\usepackage{color}
\usepackage{epstopdf}
\newcommand{\lsim}{\mathrel{\mathop{\kern 0pt \rlap
  {\raise.2ex\hbox{$<$}}}
  \lower.9ex\hbox{\kern-.190em $\sim$}}}
\newcommand{\gsim}{\mathrel{\mathop{\kern 0pt \rlap
  {\raise.2ex\hbox{$>$}}}
  \lower.9ex\hbox{\kern-.190em $\sim$}}}

\newcommand{\beq}     {\begin{equation}}
\newcommand{\eeq}     {\end{equation}}
\newcommand{\bea}     {\begin{eqnarray}}
\newcommand{\eea}     {\end{eqnarray}}

\newcommand{\gev}      {{\,\rm GeV}}

\begin{document}
\title{Cosmic positron and antiproton constraints \\on 
the gauge-Higgs Dark Matter}
\author{Kingman Cheung$^{1,2,3}$, Jeonghyeon Song$^{1}$, and
Po-Yan Tseng$^{2}$}

\affiliation{
$^1$Division of Quantum Phases \& Devices, School of Physics, 
Konkuk university, Seoul 143-701, Korea \\
$^2$Department of Physics, National Tsing Hua University, 
Hsinchu 300, Taiwan
\\
$^3$Physics Division, National Center for Theoretical Sciences,
Hsinchu 300, Taiwan
}

\date{\today}

\begin{abstract}
We calculate the cosmic ray
positron and antiproton spectra of 
a gauge-Higgs dark matter candidate in a  
warped five-dimensional
$SO(5) \times U(1)$ gauge-Higgs unification model.
The stability of the gauge-Higgs boson is guaranteed
by the $H$ parity under which only the Higgs boson
is odd at low energy.
The  4-point vertices of
$HHW^+W^-$ and $HHZZ$, allowed 
 by $H$ parity conservation,
have the same magnitude as in the standard model,
which yields efficient annihilation rate for $m_H > m_W$.
The most dominant 
annihilation channel is $H H \to W^+ W^-$ followed by the
subsequent decays of the $W$ bosons into positrons or quarks,
which undergo fragmentation into antiproton. 
Comparing with the observed positron and
antiproton spectra with the PAMALA and Fermi/LAT,
we found that 
the Higgs boson mass cannot be larger than 90 GeV,
in order not to overrun the observations.
Together with the constraint on not overclosing the Universe, the
valid range of the dark matter mass is restricted to
$70-90$ GeV.
\end{abstract}
\maketitle

\section{Introduction}
Since Lee and Weinberg put bound on a stable neutrino mass
from the relic density of the Universe in 1977\,\cite{Lee:1977ua},
various cosmological and astrophysical observations
have inspired and also constrained many theoretical models
in particle physics. 
In particular the very precise measurement 
of the cosmic microwave background radiation
in the Wilkinson Microwave Anisotropy Probe (WMAP) experiment
established the 
presence of cold dark matter (CDM) in our Universe\,\cite{wmap}.
These data are urging new physics (NP) 
beyond the standard model (SM) to 
provide the stable CDM particle
over cosmological time scale.

Recently, a number of high energy cosmic ray experiments
suggested the possibility of indirect detections
of CDM annihilation in the galactic halo.
The Payload for Antimatter Matter Exploration and 
Light-nuclei Astrophysics (PAMELA) collaboration reported
the excess of positrons in its energy spectrum 
of $10-100$ GeV over the expected high energy cosmic ray 
interaction with the interstellar medium\,\cite{pamela-e}.
More mysterious result was that the antiproton
data, on the other hand, seem to be consistent with the expected
spectrum from the 
collisions between high-energy protons in cosmic rays 
and the nuclei of hydrogen and helium atoms 
in the interstellar medium\,\cite{pamela-p}.
In addition, the Fermi/LAT\,\cite{fermi} and 
the HESS\,\cite{hess} collaborations also
reported a smooth but harder electron-flux spectrum than 
the expected background at 100--1000 GeV.

In the literature there have been great efforts
to explain these anomalous data
by the annihilation of CDM particles
in NP models,
such as the lightest supersymmetric particle\,\cite{pamela-susy},
the Kaluza-Klein particle\,\cite{pamela-KK},
and other CDM candidates\,\cite{pamela-other}.
It was shown that many CDM particle candidates such as in
supersymmetric (SUSY) models and the
universal extra dimensional 
(UED) model
require a large boost factor
to explain the observed signal.
Even though the inhomogeneity of CDM\,\cite{boost} 
or the Sommerfeld
mechanism\,\cite{sommerfeld} can help, the values of
these boost factors are limited.
Some other NP models\,\cite{pamela-other}, e.g.,
a new long range force 
in the dark sector\,\cite{ArkaniHamed:2008qn}, 
can explain the PAMELA data without extreme boost factor.

In this paper we adopt a different approach
to these astrophysical data, i.e., 
using them to constrain a NP theory, instead of
explaining them to support a theory.
The positron excess could be explained by
astrophysical sources like pulsars\,\cite{pulsar}
or supernova remnants\,\cite{supernova}.
Historically the astrophysical observations have
played the role of a strong constraint on a 
NP model.
For instance, 
the very precise WMAP measurement of the relic density
eliminates a large portion of 
the parameter space of the constrained minimal
supersymmetric SM to avoid overclosing
the Universe\,\cite{Ellis:2003cw}.

We will study the positron and antiproton energy spectra
from CDM annihilation
in  a recently proposed  
$SO(5) \times U(1)$
gauge-Higgs unification model 
based on a warped five-dimensional (5D) 
spacetime\,\cite{hosotani08,hosotani09,hosotani10}.
Here the dark matter is nothing but
the Higgs boson\,\cite{ko}, which is a part of the 
fifth component of a gauge boson field in the model.
The $H$ parity,
under which only the Higgs boson has odd parity
at low energy,
is preserved from the gauge structure of the theory.
Triple vertices such as $WWH$, $ZZH$, and $\bar{f}f H$
(here $f$ is a SM fermion)
vanish to all orders in perturbation theory\,\cite{Sakamura,ko}, thus
the Higgs boson is stable and becomes a dark matter candidate.
Instead, the 4-point vertices of $HHW^+ W^-$, $HHZZ$,
and $HHf\bar{f}$ are allowed by the $H$ parity conservation.
The Higgs boson can be thermally produced 
in the early Universe via
$WW,ZZ,f\bar f \to H H$, of which the rate is 
highly predictive with essentially one 
free parameter, the Higgs boson mass $m_H$.

In Ref.\,\cite{ko},  it was shown that 
the Higgs boson mass needs to be
at 70 GeV in order to explain the WMAP data. 
When $m_H < 70$ GeV, the total annihilation
rate is very small 
because the kinematically allowed $2\to 2$ processes
into light fermions are suppressed by small Yukawa couplings.
The corresponding relic density becomes too big,
which overcloses the Universe.
If $m_H>70$ GeV, the cross section
of $HH\to WW^{(*)}$ becomes very large,
since the magnitude of $HHWW$ coupling is the same as in the SM.
The corresponding relic density is too low.
Nevertheless, this is not ruled out. 
It is possible that
the dark matter can be produced nonthermally from, e.g., other
quasi-stable Kaluza-Klein states or other topological objects. 
In this work, we assume that the thermal source is not the only source
of relic dark matter, and so Higgs boson mass can be larger than 70 GeV.
The present upper bound on the Higgs mass comes 
from the unitarity requirement, which limits
$m_H<{\mathcal O}(1)$ TeV\,\cite{haba}.

We note that this large Higgs mass yields too efficient 
annihilation into $W^+ W^-$ and possibly $ZZ$,
which leads to potentially large positron and antiproton signals.
Using the observed positron and antiproton spectra in the PAMELA
and Fermi/LAT experiments,
we can set the upper limit on the Higgs boson mass.
On the other hand, the collider signal for this model is way too
small for detection\,\cite{CS}.

In this work, we study the positron and antiproton spectra from
annihilation of the gauge-Higgs dark matter in the halo.
We use the cosmic ray propagation code
Galprop\,\cite{galprop} to calculate the propagation of the positron
and antiproton from the halo to the Earth, and compare with the
spectra measured by PAMELA\,\cite{pamela-e,pamela-p}.  The most
dominant annihilation channel for $m_H \ge 70$ GeV 
is $H H \to W^+
W^-$, followed by the subsequent decays of the $W$ bosons into
positrons or quarks.
The quarks undergo fragmentation into antiproton.
Here we do not attempt to explain the anomaly observed by PAMELA.
Instead we use them as the constraints on the model:
the resulting spectra obtained from the CDM
annihilation should not exceed the ones measured by PAMELA.

It is well known that the process $H H \to W^+ W^-$ 
grows with
the center-of-mass (c.m.) energy $\sqrt{s}$. 
Naively, the longitudinal 
polarization of the $W$ boson behaves like $p^\mu/m_W$
when $\sqrt{s} \gg m_W$.  Therefore,
we expect 
the scattering amplitude squared grows as $s^2/m_W^4$ before any UV physics comes in 
to unitarize the theory.\footnote{In this gauge-Higgs model 
based on five-dimensional
spacetime, it is the
Kaluza-Klein states (of order TeV) of the gauge bosons which 
unitarize the scattering amplitude.}
For the CDM annihilation in the halo
where $v\approx 10^{-3}$, 
the c.m. energy is just roughly $2 m_H$.  
Thus, we
expect the annihilation rate will grow with the Higgs boson mass, 
as rapidly as $m_H^2/m_W^2$, so long as $m_H < 1$ TeV.  
We will show that
when the Higgs boson mass is 90 GeV or above, the resulting
positron spectrum is already well above that measured by PAMELA.
Similar conclusion holds true for the antiproton spectrum.  
We can, therefore,
conclude that
the Higgs boson mass cannot be larger than about 90 GeV
in this gauge-Higgs model.  This is the main result of our work.

The organization of the paper is as follows.
In the next section, we briefly review the model and describe
the effective interactions used in this calculation.
We give details about the calculation of the
positron spectrum and antiproton spectrum 
in Secs.\,\ref{sec:positron} and \ref{sec:antiproton}, respectively.
We also perform the comparison with the measured spectra. 
We conclude in Sec.\,\ref{sec:conclusions}. 

\section{The Effective Interactions 
in the $SO(5) \times U(1)$ gauge-Higgs model}
\label{sec:model}

The model under consideration is a $SO(5) \times U(1)$ gauge-Higgs
unification model 
in the 5D Randall-Sundrum warped space\,\cite{hosotani08}.
The Higgs boson is the fluctuation mode of the
Aharonov-Bohm phase $\hat{\theta}_H$ along the fifth dimension\,\cite{ab-phase}, 
i.e., $\hat{\theta}_H = \theta_H + H(x) /f_H$.
The four-dimensional 
effective Lagrangian of the Higgs boson
is
\begin{equation}
\label{eq:effective:V}
{\cal L} = V_{\rm eff}(\hat{\theta}_H)
- m_W^2(\hat{\theta}_H) W^+_\mu W^{-\mu} 
- \frac{1}{2}m_Z^2(\hat{\theta}_H) Z_\mu Z^\mu
- \sum_{f} m_{f}(\hat{\theta}_H) \bar \psi_f \psi_f  \;,
\end{equation}
where the mass functions are
\begin{eqnarray}
\label{eq:mw:mz}
m_W(\hat{\theta}_H) =
\frac{1}{2} g f_H \sin \hat{\theta}_H
, \quad
m_Z(\hat{\theta}_H) =
\frac{1}{2} g_Z f_H \sin \hat{\theta}_H,
\quad
m_f(\hat{\theta}_H) = y_f f_H \sin \hat{\theta}_H \;.
\end{eqnarray}
Here $g$ is the weak gauge coupling
and $g_Z = g/\cos\theta_W$.

The effective potential $V_{\rm eff}(\hat{\theta}_H)$
of the Higgs boson is
generated at one loop level. 
It is finite and cutoff 
independent.
As shown in Ref.\,\cite{hosotani08},
the large contribution of 5D top quark field
changes the global minimum of 
$V_{\rm eff}(\hat{\theta}_H)$
into $\hat{\theta}_H = \pm\pi/2$:
the $W$, $Z$ gauge bosons as well as the SM fermions
acquire their masses (see Eq.(\ref{eq:mw:mz}))
and thus the electroweak symmetry is broken dynamically.

In this model the global minimum at $\hat{\theta}_H = \pm\pi/2$
dynamically generates
a new $H$ parity,
under which the Higgs boson has odd parity
while all the other SM particles have even parity.
This $H$-parity prohibits triple vertices of the Higgs boson
with the SM particles, such as $H W^+ W^-$,
$HZZ$ and $H f \bar{f}$:
it preserves the stability of the Higgs boson
so that the Higgs boson can be 
a CDM candidate.

At low energy this model has two parameters,
$f_H$ and $m_H$.
The parameter
$f_H$ is determined by the observed $m_W$ and $m_Z$,
i.e., $f_H \approx 246\gev$.
The value of $m_H$ is, in principle,
determined if the whole
matter content in the model is fixed in detail.
Without \textit{a priori} knowledge of UV physics,
we treat $m_H$ as a free parameter.

Because of the absence of the triple vertices 
of the Higgs boson with the SM particles,
we do not have significant constraint
on $m_H$ from collider physics phenomenology.
Instead
the observed relic density of CDM 
in the WMAP experiment
can provide a meaningful one\,\cite{ko}.
If the Higgs boson mass becomes heavier,
their annihilation into $W^+ W^-$ and possibly into $ZZ$
are kinematically accessible, leading to
smaller relic density of the Universe.
On the other hand, lighter $m_H$ yields too small
annihilation cross section, which is excluded
as overclosing the Universe.
The Higgs boson mass $m_H =70\gev$
can explain the observed relic density.
We employ the WMAP data
as not overclosing 
the universe: the Higgs boson mass can be heavier than $70\gev$.

The effective interactions used in this work are
\begin{equation}
{\cal L} = \frac{1}{8}g_Z^2 H^2 Z_\mu Z^\mu
+ \frac{1}{4} g^2 H^2 W^+_\mu W^{-\mu}
+ \sum_f \frac{m_f}{2 f_H^2} H^2 \bar \psi_f \psi_f.
\end{equation}
Note that the $HHW^+W^-$ and $HHZZ$ vertices
have the same couplings with the SM ones
except for the opposite sign.
Since the annihilation cross section
of $HH\to W^+W^-,ZZ$ increases with the c.m. energy $\sqrt{s}$
(or $m_H$ when the relative velocity of two $H$'s is
very small),
heavier Higgs boson could leave too much excess of positrons
and antiprotons from $W$ or $Z$ decays.

\section{Positron Spectrum}
\label{sec:positron}

The dominant process for the cosmic ray positrons
from the Higgs boson annihilation is
the 
leptonic decay of the $W^+$ boson:
\begin{equation}
H H \to W^+ W^- \to e^+ + \nu_e + X \,.
\end{equation}
The next dominant process is
$HH \to Z Z \to e^+ + X$.
Positrons can also come from the hadrons, which are
the fragmentation products of the quarks from $W$ decays.
These positrons are much softer than those coming directly
from the $W$ boson decay. We shall ignore these soft positrons.
There are also other processes $H H \to f \bar f$ (for $f=b,c$)
to produce positrons in the subsequent decays of the 
fragmentation products,
but they are certainly subleading because of 
the small Yukawa couplings compared with the gauge coupling. 

A back-of-envelope calculation shows that the annihilation rate is given
by
\begin{eqnarray}
\langle \sigma v \rangle_{HH \to WW} &\equiv&
\sigma(H H \to W^+ W^-) \, (2\beta_H) = \frac{g^4 \beta_W}{32 \pi s}\,
\left( 3 - \frac{s}{m_W^2} + \frac{s^2}{4 m_W^4} \right ) \;, \\
\langle \sigma v \rangle_{HH \to ZZ} &\equiv&
\sigma(H H \to ZZ ) \, (2\beta_H) = \frac{g_z^4 \beta_Z}{64 \pi s}\,
\left( 3 - \frac{s}{m_Z^2} + \frac{s^2}{4 m_Z^4} \right ) \;,
\end{eqnarray}
where $2\beta_H = 2 \sqrt{ 1 - 4m_H^2/s}$ is 
the relative velocity of the two non-relativistic 
incoming Higgs bosons in their c.m. frame, 
and $\beta_{W,Z} = \sqrt{ 1 - 4m_{W,Z}^2/s}$.  It is easy to see that
the annihilation rate grows as $s/m_{W,Z}^2$. 

The positron flux observed at the Earth is given by
\begin{equation}
 \Phi_{e^+} (E) = \frac{ v_{e^+} } { 4 \pi} \, f_{e^+} (E) \;,
 \label{semiflux} 
\end{equation}
with $v_{e^+}$ is close to the velocity of light $c$.
The function 
$f_{e^+} (E)$ satisfies the diffusion equation of 
\begin{equation}
\frac{\partial f}{\partial t} - K(E)  \nabla^2 f 
- \frac{\partial}{\partial E} \left( b(E) f \right ) = Q \;,
\end{equation}
where the diffusion coefficient is
$K(E) = K_0(E/{\rm GeV})^\delta$ and 
the energy loss coefficient is 
$b(E) = E^2/ ({\rm GeV} \times \tau_E)$
with $\tau_E = 10^{16}$ sec. 
The source term $Q$ due to the annihilation is 
\begin{equation}
Q_{\rm ann} = \eta \left( \frac{\rho_{\rm CDM} }{M_{\rm CDM}} \right )^2 
\, \sum \langle \sigma v \rangle_{e^+} \, \frac{d N_{e^+}}{ d E_{e^+} } \;,
\end{equation}
where $\eta = 1/2 $ for identical scalar DM particle in the
initial state.
The summation is over all possible channels that can produce positrons in 
the final state, and $dN_{e^+}/dE_{e^+}$ denotes the spectrum of the positron
energy per annihilation in that particular channel.

In our analysis, the source term is given by
\begin{equation}
Q_{\rm ann} = \frac{1}{2} \left( 
\frac{\rho_{\rm CDM} }{M_{\rm CDM}} \right )^2 
\, \left[ 
   \langle \sigma v \rangle_{HH\to W W} \, \frac{d N^{WW}_{e^+}}{ d E_{e^+} } 
+ \langle \sigma v \rangle_{HH\to ZZ } \, \frac{d N^{ZZ}_{e^+}}{ d E_{e^+} } 
\right ] \;,
\end{equation}
where the normalizations of $N^{WW}_{e^+}$ and $N^{ZZ}_{e^+}$
are
\beq
\int \frac{d N^{WW}_{e^+} }{d x} dx =B(W^+ \to e^+ \nu_e),
\quad
\int \frac{d N^{ZZ}_{e^+} }{d x} dx = 2 \times B(Z  \to e^+ e^-).
\eeq
We first calculate the energy spectrum of the positron in the 
$W^+$ or $Z$ rest frame, then boost it to the c.m. frame of the
$HH$ system. 
In the calculation we include all the off-shell effects of the 
$W^+$ and $W^-$ bosons and in both $Z$ bosons.
We then put the source term into Galprop\,\cite{galprop} 
to solve the diffusion equation.

Note that the contribution from $HH \to ZZ$ channel 
is subleading 
in our calculation.  
It only accounts for about 1\% and 4\% 
of the contribution from $HH\to W^+ W^-$
channel for 
$m_H=82$ and 90 GeV, respectively.
For $m_H=100$ GeV,
the $ZZ$ channel contribution is as large as 20\% of the $WW$ one.  
It is easy to understand: for $m_H$ below 90 GeV 
the $ZZ$ channel is below the production
threshold and thus one of the $Z$ bosons has to be off-shell, 
while
at $m_H =100$ GeV both $Z$ bosons are already on shell.

\begin{figure}[t!]
\centering
\includegraphics[width=5in]{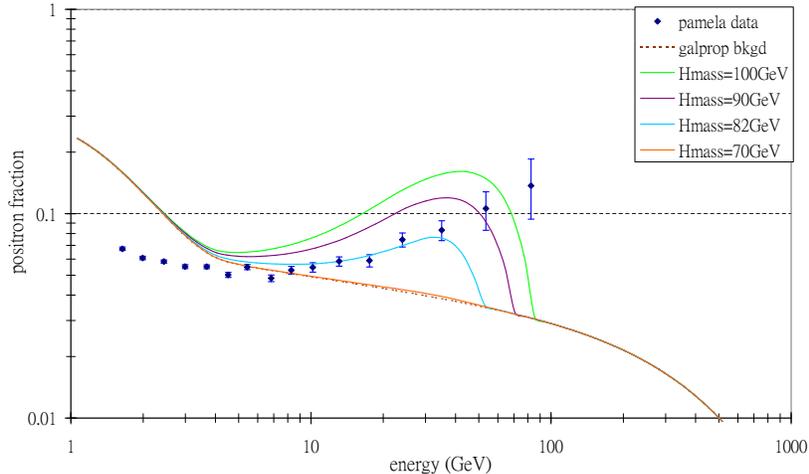}
\caption{\small \label{positron_spectrum}
The positron-fraction spectrum predicted for the gauge-Higgs dark matter model
for $m_H=70, 82, 90, 100$ GeV, assuming that it accounts for all the
dark matter of the Universe.  The PAMELA data are also shown.
}
\end{figure}

Figure \ref{positron_spectrum} shows
the positron spectra  
for $m_H = 70, 82, 90$, and 100 GeV
with the measured spectrum
of PAMELA\,\cite{pamela-e}.
For $m_H =70\gev$,
the gauge-Higgs CDM annihilation cannot explain the
rising feature in the positron spectrum.
The annihilation rate is too low
since the $W^+ W^-$ channel is not kinematically open yet.
When the Higgs boson mass
is above $m_W$, 
the annihilation rate increases quickly and can explain
part of the rise-up in the spectrum, 
as shown by the $m_H=82$ GeV curve.
Yet, it still cannot explain the two highest energy points 
because the
mass of the Higgs boson is not heavy enough.  
Once  the Higgs boson mass  rises to
$90-100$ GeV, it passes through the second highest energy point, but 
the annihilation rate increases far more than the lower part of the
spectrum allowed. 
Just by visual checking, $m_H$ larger than 90 GeV 
is easily ruled out
by the measured spectrum.  

\begin{figure}[t!]
\centering
\includegraphics[width=5in]{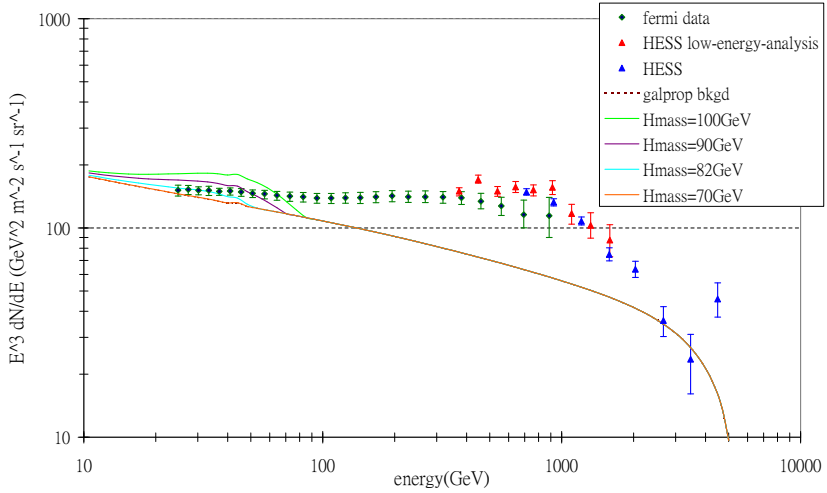}
\caption{\small \label{total_spectrum}
Total flux of electron and positron predicted for the 
gauge-Higgs dark matter model for $m_H = 70, 82, 90, 100$ GeV.
Data from Fermi/LAT and HESS are shown.}
\end{figure}

We also show the total electron and positron flux 
in Fig. \ref{total_spectrum}.
Our results are compared with the
measurement of the cosmic ray 
$e^+ + e^-$ spectra from 20 GeV to 1 TeV 
with the Fermi LAT\,\cite{fermi}, and from 340 GeV
with the HESS\,\cite{hess}.
These cosmic ray electron spectra
are distributed over very high energy up to 1 TeV.
Unless the CDM mass is very heavy
about $800-1000$ GeV,
the whole energy spectrum cannot be explained.
Instead of explaining high energy spectrum,
we focus on the region of $20 \gev \lsim E \lsim 100 \gev$.
The case of $m_H=82\gev$ is marginally allowed by the
data while the $m_H =90\gev$ case
already outruns the observation. 

\section{Antiproton Spectrum}
\label{sec:antiproton}

Similar to the treatment for positron flux, the antiproton
flux can be obtained by solving the diffusion equation
with an appropriate source term for the input antiproton spectrum:
\begin{equation}
Q_{\rm ann} = \eta \left( \frac{\rho_{\rm dm} }{M_{\rm dm}} \right )^2 
\, \sum \langle \sigma v \rangle_{\bar p} \, \frac{d N_{\bar p}}{ d T_{\bar p} }
 \;,
\end{equation}
where $\eta =1/2$, and $T_{\bar p}$ is 
the kinetic energy of the antiproton
which is conventionally used instead of the total energy. 
We solve the diffusion equation using Galprop\,\cite{galprop}.

In our case, the dominant contribution 
to the cosmic ray antiproton production comes from
\begin{equation}
H H \to W^+ W^- \to (q \bar q') ( q\bar q') \to \bar p + X \;,
\end{equation}
followed by $HH \to ZZ \to (q \bar q) (q' \bar q') \to \bar p +X$.
In the last step, 
we adopt a publicly available
code\,\cite{kniehl} to calculate the fragmentation function $D_{q\to h}(z)$ 
for any quark $q$ into hadrons $h$, e.g., $p,\bar p, \pi$.  
The fragmentation
function is then convoluted with energy spectrum $d N/ dE$ 
of the light quark to obtain the energy spectrum of the antiproton
$d N /d E_{\bar p}$.  
The next contribution comes from 
$HH \to b \bar b \to \bar p + X$.
Since the annihilation is
smaller by two orders of magnitude, 
we ignore this and the other subleading
contributions.  The source term $dN /d T_{\bar p}$ is then 
implemented into
Galprop to calculate the propagation from the halo to the Earth.

\begin{figure}[t!]
\centering
\includegraphics[width=5in]{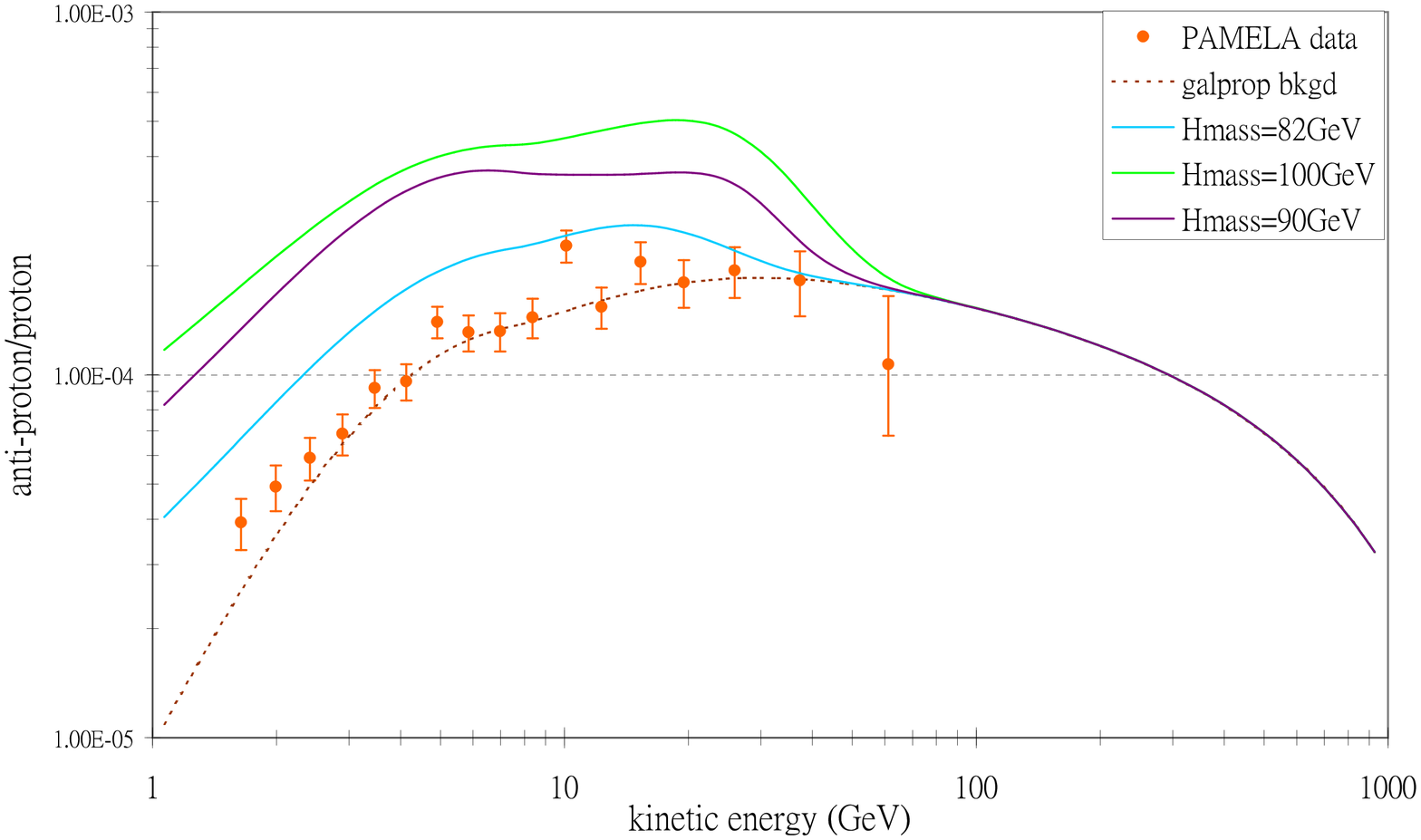}
\caption{\small \label{antiproton}
The fraction of the antiproton $\bar p/(p+\bar p$) 
versus the kinetic energy of the antiproton predicted for 
the gauge-Higgs dark matter model for $m_H= 82,90,100$ GeV.
The antiproton data of PAMELA are also shown.
}
\end{figure}

The resulting antiproton fraction as a function of its
energy is shown in Fig.~\ref{antiproton}.  
Three theoretical curves are for $m_H= 82, 90, 100$ GeV,
compared with the observed PAMELA data.  
Note that in this figure we only include the leading contribution
from $HH \to W^+ W^-$ because the next subleading contribution
only accounts for less than 5\% of the leading one for $m_H \le 90$ GeV.
We can see that the $m_H=82$ (just above the $WW$ threshold) curve 
is barely consistent
with the data. On the other hand, the $m_H=90$ and 100 GeV curves are 
obviously over the data. 

Since the theoretical background curve contains 
relatively high
uncertainties due to different models used in Galprop, 
we do not perform any confidence-level exclusion analysis. 
Instead, we only use a visual inspection.  Very conservatively,
we conclude that the $m_H=90$ GeV or above is excluded by 
the antiproton spectrum.  
A similar conclusion can be drawn from the positron spectrum. 
Overall, $m_H = 90$ GeV or above is strongly disfavored  by the 
data


\section{Conclusions}
\label{sec:conclusions}

In the $SO(5) \times U(1)$ gauge-Higgs unification model
based on the Randall-Sundrum warped spacetime,
the Higgs boson is a cold dark matter candidate.
Its stability over the cosmological time scale is
guaranteed by the $H$ parity 
under which only the Higgs boson 
has negative parity at low energy.
The triple vertices of $HWW$, $HZZ$ and $H\bar{f}f$
vanish.
However the 4-point vertices of $HHWW$, $HHZZ$,
and $HHf\bar{f}$ are allowed.

Interpreting the WMAP data
on the relic density as not overclosing the Universe,
the Higgs boson mass is constrained from below
as $m_H\ge 70\gev$. 
Noting that the heavy Higgs boson generates
more efficient annihilation into $W^+ W^-$ and $ZZ$, 
we study the cosmic ray positron and antiproton spectra
and compare them with the PAMELA and Fermi/LAT observations.
We do not aim at explaining the up-rising positron
spectrum observed with the PAMELA.
Instead we use the data to constrain the model.
The Higgs boson annihilation rate 
is shown to increase with increasing
Higgs boson mass.
The $m_H=82$ GeV is marginally consistent with 
both the observed  positron and antiproton spectra.
However, the
Higgs boson mass of 90 GeV or more is obviously ruled out 
by the observed
data.   

Our purpose is to constrain the 
$SO(5) \times U(1)$ gauge-Higgs unification model
by the PAMELA and Fermi/LAT data,
which turned out to be very significant.
The present limit
on the model comes from the unitarity requirement, which limits
the Higgs boson to be less than about ${\mathcal O}(1)$ TeV\,\cite{haba}.
We show explicitly in this paper that the Higgs boson mass 
cannot be larger than 90 GeV, otherwise it 
overruns the
observed positron and antiproton spectra of PAMELA
and Fermi/LAT.
Together with the constraint on the closure of the Universe, 
the gauge-Higgs boson mass as  CDM 
is now restricted to $70-90$ GeV.

\section*{Acknowledgments}
We thank Professor Hosotani for useful discussion on the $H$ parity.
The work was supported in parts by the NSC of Taiwan under 
Grant Nos. 96-2628-M-007-002-MY3 and the
WCU program through the KOSEF funded by the MEST
(R31-2008-000-10057-0).



\begin{thebibliography}{99}

\bibitem{Lee:1977ua}
  B.~W.~Lee and S.~Weinberg,
  Phys.\ Rev.\ Lett.\  {\bf 39}, 165 (1977).

\bibitem{wmap}
J.~Dunkley {\it et al.}  [WMAP Collaboration],
  Astrophys.\ J.\ Suppl.\  {\bf 180}, 306 (2009).


\bibitem{pamela-e}
 O.~Adriani {\it et al.}  [PAMELA Collaboration],
  Nature {\bf 458}, 607 (2009).

\bibitem{pamela-p}
O.~Adriani {\it et al.},
  Phys.\ Rev.\ Lett.\  {\bf 102}, 051101 (2009).

\bibitem{fermi}
  A.~A.~Abdo {\it et al.}  [The Fermi/LAT Collaboration],
  Phys.\ Rev.\ Lett.\  {\bf 102}, 181101 (2009).

\bibitem{hess}
  F.~Aharonian {\it et al.}  [H.E.S.S. Collaboration],
  Astron.\ Astrophys.\  {\bf 508}, 561 (2009).

\bibitem{pamela-susy}
See e.g., 
L.~Bergstrom, T.~Bringmann and J.~Edsjo,
  Phys.\ Rev.\  D {\bf 78}, 103520 (2008);
K.~Ishiwata, S.~Matsumoto and T.~Moroi,
  Phys.\ Lett.\  B {\bf 675}, 446 (2009);
P.~Grajek, G.~Kane, D.~Phalen, A.~Pierce and S.~Watson,
  Phys.\ Rev.\  D {\bf 79}, 043506 (2009);
J.~H.~Huh, J.~E.~Kim and B.~Kyae,
  Phys.\ Rev.\  D {\bf 79}, 063529 (2009);





\bibitem{pamela-KK}

See e.g., 
D.~Hooper and K.~M.~Zurek,
  Phys.\ Rev.\  D {\bf 79}, 103529 (2009);
Y.~Bai and Z.~Han,
  Phys.\ Rev.\  D {\bf 79} (2009) 095023;
S.~C.~Park and J.~Shu,
  Phys.\ Rev.\  D {\bf 79}, 091702 (2009).



\bibitem{pamela-other}
See e.g., 
 A.~Ibarra and D.~Tran,
  JCAP {\bf 0902}, 021 (2009);
P.~J.~Fox and E.~Poppitz,
  Phys.\ Rev.\  D {\bf 79}, 083528 (2009);
Y.~Nomura and J.~Thaler,
  Phys.\ Rev.\  D {\bf 79}, 075008 (2009);
R.~Harnik and G.~D.~Kribs,
  Phys.\ Rev.\  D {\bf 79}, 095007 (2009);
I.~Cholis, D.~P.~Finkbeiner, L.~Goodenough and N.~Weiner,
  JCAP {\bf 0912}, 007 (2009);
D.~Feldman, Z.~Liu and P.~Nath,
  Phys.\ Rev.\  D {\bf 79}, 063509 (2009);
C.~R.~Chen, F.~Takahashi and T.~T.~Yanagida,
  Phys.\ Lett.\  B {\bf 671}, 71 (2009);
M.~Cirelli, M.~Kadastik, M.~Raidal and A.~Strumia,
  Nucl.\ Phys.\  B {\bf 813}, 1 (2009);
V.~Barger, W.~Y.~Keung, D.~Marfatia and G.~Shaughnessy,
  Phys.\ Lett.\  B {\bf 672}, 141 (2009);
E.~J.~Chun and J.~C.~Park,
  JCAP {\bf 0902}, 026 (2009).




\bibitem{boost}
J.~Diemand, M.~Kuhlen, P.~Madau, M.~Zemp, B.~Moore, D.~Potter and J.~Stadel,
  Nature {\bf 454}, 735 (2008).



\bibitem{sommerfeld}
M.~Lattanzi and J.~I.~Silk,
  Phys.\ Rev.\  D {\bf 79}, 083523 (2009);
J.~Bovy,
  Phys.\ Rev.\  D {\bf 79}, 083539 (2009).



\bibitem{ArkaniHamed:2008qn}
  N.~Arkani-Hamed, D.~P.~Finkbeiner, T.~R.~Slatyer and N.~Weiner,
  Phys.\ Rev.\  D {\bf 79}, 015014 (2009).


\bibitem{pulsar}
D.~Malyshev, I.~Cholis and J.~Gelfand,
  Phys.\ Rev.\  D {\bf 80}, 063005 (2009);
D.~Hooper, P.~Blasi and P.~D.~Serpico,
  JCAP {\bf 0901}, 025 (2009);
 H.~Yuksel, M.~D.~Kistler and T.~Stanev,
  Phys.\ Rev.\ Lett.\  {\bf 103}, 051101 (2009);
S.~Profumo,
  arXiv:0812.4457 [astro-ph].

\bibitem{supernova}
N.~J.~Shaviv, E.~Nakar and T.~Piran,
  Phys.\ Rev.\ Lett.\  {\bf 103}, 111302 (2009).

\bibitem{Ellis:2003cw}
  J.~R.~Ellis, K.~A.~Olive, Y.~Santoso and V.~C.~Spanos,
  Phys.\ Lett.\  B {\bf 565}, 176 (2003).

\bibitem{hosotani08}
Y.~Hosotani, K.~Oda, T.~Ohnuma and Y.~Sakamura,
  Phys.\ Rev.\  D {\bf 78}, 096002 (2008)
  [Erratum-ibid.\  D {\bf 79}, 079902 (2009)].

\bibitem{hosotani09}
Y.~Hosotani and Y.~Kobayashi,
  Phys.\ Lett.\  B {\bf 674}, 192 (2009).

\bibitem{hosotani10}
  Y.~Hosotani,
  arXiv:1003.3129 [hep-ph].


\bibitem{ko}
Y.~Hosotani, P.~Ko and M.~Tanaka,
  Phys.\ Lett.\  B {\bf 680}, 179 (2009).

\bibitem{Sakamura}
  Y.~Sakamura,
  Phys.\ Rev.\  D {\bf 76}, 065002 (2007).

\bibitem{haba}
 N.~Haba, Y.~Sakamura and T.~Yamashita,
  JHEP {\bf 1003}, 069 (2010).

\bibitem{CS}
K.~Cheung and J.~Song,
 Phys. Rev. \ D {\bf 81}, 097703 (2010).

\bibitem{galprop}
A.~W.~Strong, I.~V.~Moskalenko, T.~A.~Porter, G.~Johannesson, 
E.~Orlando and S.~W.~Digel,
  arXiv:0907.0559 [astro-ph.HE].

\bibitem{ab-phase}
Y.~Hosotani,
  Phys.\ Lett.\  B {\bf 126}, 309 (1983);
A.~T.~Davies and A.~McLachlan,
  Phys.\ Lett.\  B {\bf 200}, 305 (1988);
Y.~Hosotani,
  Annals Phys.\  {\bf 190}, 233 (1989).

%
%

\bibitem{kniehl}
S.~Albino, B.~A.~Kniehl and G.~Kramer,
  Nucl.\ Phys.\  B {\bf 725}, 181 (2005).


\end{thebibliography}
\end{document}